\documentclass{article}
\usepackage{amsmath}
\begin{document}  
 
\title{Magnetic Surfaces in Stationary Axisymmetric General Relativity}  
\author{L. Fern\'andez-Jambrina\\ 
Departamento de Geometr\'{\i}a y Topolog\'{\i}a,\\
Facultad de Ciencias Matem\'aticas,\\
Universidad Complutense de Madrid\\
E-28040-Madrid, Spain}   
\maketitle 
\begin{abstract} 
In this paper a new method
is derived for constructing electromagnetic surface sources for stationary
axisymmetric electrovac spacetimes endowed with non-smooth or even discontinuous 
 Ernst potentials.
This can be viewed as a generalization of some classical potential theory
results, since lack of continuity of the potential is related to dipole density
and lack of smoothness, to monopole density. In particular this approach is
useful for constructing the dipole source for the magnetic field. This formalism
involves solving a linear elliptic differential equation with boundary
conditions at infinity. As an example, two different models of surface densities
for the Kerr-Newman electrovac spacetime are derived.  
\end{abstract}   

\section{Introduction}

One of the major challenges of general relativity is the description of a
compact rotating material gravitational source and the vacuum surrounding it. Its
astrophysical interest is clear, since it would be useful for modelling
relativistic rotating stars and galaxies. Nevertheless, there are no known exact
solutions for this problem so far. If we allow the vacuum spacetime to have
isometries such as stationarity and axial symmetry, the number of solutions of
the Einstein equations that can be obtained by means of generation techniques
(cfr. for instance \cite{Corn}) is huge. On the contrary there is only a limited
number of exact solutions, sharing the same symmetries, that could be regarded as
inner material sources (perfect fluids with a physically reasonable equation of
state, for instance) and none of these has been smoothly matched to an
asymptotically flat vacuum spacetime (cfr. \cite{Esc} for a recent review on the
subject).

 In section \ref{match} an outline of the mathematical problem of
matching spacetimes is provided before dealing with thin layers, that is,
two-dimensional sources that could be interpreted as limits of physical
configurations in which one of the characteristic lengths of the source can be
neglected when compared with the others. In order to shed light on the method
for constructing dipole distributions, a reminder of some formulae of potential
theory is given in section \ref{potential}. These expressions are generalized to
stationary axisymmetric electrovacuum spacetimes in section
\ref{relat}. The role of the scalar
potential is taken on by the complex Ernst potential \cite{Ernst}, which was
introduced for solving the Einstein-Maxwell system with a two-dimensional group
of isometries on the underlying manifold. The lack of continuity of this
potential will give rise to a source for the fields consisting of monopole and
dipole layers. The results will be applied to the Kerr-Newman metric in section
\ref{example} and compared with the ones achieved by Israel in \cite{is1}. A
brief discussion is provided.

\section{The matching problem} \label{match}

Our aim is the mathematical description of a relativistic compact object and
the gravitational field in the vacuum that surrounds it. Therefore it is
convenient to consider two Lorentzian manifolds $(M_+,g_+)$, $(M_-,g_-)$, 
respectively corresponding to the outer vacuum and the inner material source, 
whose matter content satisfies the Einstein equations,

\begin{equation}
Ricci(g_\pm)-\frac{1}{2}\,R\,g_\pm=8\,\pi\,T_\pm.
\end{equation}

The stress tensor in $(M_+,g_+)$ can be either electromagnetic or vacuum
whereas in $(M_-,g_-)$ it can be the one corresponding to an isentropic
(charged) perfect fluid.

As a simplification we shall allow both spacetimes to have isometries and
restrict ourselves to stationary axisymmetric spacetimes. The Killing fields
that implement these isometries, $\xi$, $\eta$, respectively the generators of
the stationary and axial symmetry, must fulfil certain conditions,

\begin{equation}
g(\xi,\xi)<0\ \ \ \ g(\eta,\eta)\ge 0\ \ \ \ [\xi,\eta]=0.
\end{equation}
and $\eta$ must have closed orbits. 

The symmetry axis will be then defined by the set of events where the
axial Killing field is a null vector,

\begin{equation}
\Delta=g(\eta,\eta)=0,
\end{equation}
and in order to avoid conical singularities on it, we have to impose a
regularity condition, namely, \cite{Kramer}

\begin{equation}
\frac{g({\rm grad}\,\Delta,{\rm grad}\,\Delta)}{4\,\Delta}\rightarrow 1,
\end{equation}
on approaching the axis. One can also impose that test particles moving along
the axis do not experience gravitational or electromagnetic forces that push
them away from it. This means \cite{Bergh} that in a chart where the coordinates
$\{x^0,x^1,x^2,x^3\}$ are required to satisfy that $\xi=\partial_{x^0}$,
$\eta=\partial_{x^1}$, $x^2=\sqrt{g(\eta,\eta)}$ and that the $x^2=constant$ and
$x^3=constant$ hypersurfaces are orthogonal to each other, the connection
coefficients $\Gamma^2_{\mu\nu}$ vanish on the axis for $\mu\ne2\ne\nu$.
Furthermore the same restriction on charged test particles implies that the
partial derivatives of the electromagnetic connection along the $x^2$ direction
also vanish on the axis.

We
shall also require that not only the metric but also the electromagnetic
curvature, $F$, and its four-dimensional Hodge dual,  $^4*F$, and the
4-velocity, the density and the pressure of the fluid have zero Lie-derivatives
along the Killing fields.

In order to model compact objects such as stars, it is expected that the
gravitational field will decay to zero at infinity and therefore it is
reasonable to require that the $(M_+,g_+)$ be asymptotically flat in a convenient
set of coordinates. We shall extend on this subject in the following sections.

Since we are dealing with rotating objects, nonstaticity has to be achieved.
Otherwise we would have the Schwarzchild solution. Therefore {\it every} 
timelike Killing field $\zeta$ must be non-surface-forming,

\begin{equation}
\zeta^\flat\wedge d\zeta^\flat\ne 0,
\end{equation}
where $\flat$  denotes the isomorphism between the tangent and the cotangent
bundle induced by the metric structure.

If both spacetimes are to be matched, then there must be in each of them a closed
3-dimensional timelike surface $\Sigma_\pm$ that can be imbedded in both
$(M_+,g_+)$ and $(M_-,g_-)$ as a submanifold $(\Sigma_\pm,i^*_\pm g_\pm)$

\begin{equation}
i_\pm:\Sigma_\pm\hookrightarrow M_\pm.
\end{equation}

According to Darmois' junction conditions \cite{Darmois} both spacetimes are
suitable for matching if the induced metric and the extrinsic curvature on the
hypersurfaces are continuous,

\begin{equation}
[i^*g]=0\label{induced}
\end{equation}

\begin{equation}
[K]=0\label{extrinsic},
\end{equation}
after identifying events on $\Sigma_+$ and $\Sigma_-$
by means of an isometry. The difference between the
values of a quantity on $\Sigma_+$ and $\Sigma_-$ has been denoted by a square
bracket.

If the electromagnetic field $F$ is non-zero, for the sake of continuity
it must fulfil the condition \cite{is1},

\begin{equation}
[i^*F(n)]=0\label{optic},
\end{equation}
on the matching hypersurface, whose outer unit normal is denoted by $n$.

As Israel has shown \cite{ISR}, whenever the equation (\ref{extrinsic}) is not
satisfied, the jump in the extrinsic curvature $K$ reveals the presence of an
energy-momentum surface density, $S$, on $\Sigma$,

\begin{equation}
S=\frac{1}{8\,\pi}\,([K]-Tr[K]\,i^*g)\label{crust},
\end{equation}
and in this case the inner source is surrounded by a `crust' of matter.

In the same fashion, there is a non-zero electromagnetic surface current
$j$ on $\Sigma$, given by the expression,

\begin{equation}
[i^*F(n)]=-4\,\pi\,j\label{optic2},
\end{equation}
if the continuity condition (\ref{optic}) on $F$ is not satisfied.

Thin layers are a special case of non-smooth matching. If we try to match two
vacuum  spacetimes across a common hypersurface so that the induced metric on
both sides fulfils (\ref{induced}), then the only possibility of having matter
content in the global manifold is given by (\ref{crust}), (\ref{optic2}). In
this case the `crust' is all we have.

We would be interested then in drawing physical information about the material
source that it is located on the hypersurface, such as the densities of the
physical quantities. In particular dipole densities will be the main concern in
the following sections.

\section{Classical dipole densities}\label{potential}

Before describing our approach to calculate dipole surface densities in the
relativistic situation, it will be useful to recall some results of the
classical potential theory in order to establish a comparison when dealing with
general relativity.

Let us consider a classical field which can be derived from a scalar potential,
$V$, that satisfies the Laplace equation. Assume that the field is generated by
a source located on a closed surface $S$ (a sheet of monopoles and dipoles). Then
the density of the monopole layer, $\sigma_Q$, and the density, $\sigma_M$, of
the projection of the moment in the
$z$ direction (its unit vector will be represented by $u_z$), 

\begin{subequations}
\begin{equation} 
\sigma_Q=-\frac{1}{4\,\pi}\,\left[\frac{dV}{dn}\right]\label{mono}
\end{equation}

\begin{equation} 
\sigma_M=\frac{1}{4\,\pi}\,\left\{n\cdot
u_z\,[V]-z\,\left[\frac{dV}{dn}\right]\right\},\label{clasden}  
\end{equation}
\end{subequations}
can be calculated in terms of the discontinuities (denoted by square brackets) of
the potential and its derivative along the outer unit normal, $n$, to the
surface. 

The latter expression
can be easily obtained applying the Green identity to the flat space $\Omega$,
excluding the source, 

\begin{eqnarray}
 0=\int_\Omega\,d^3x\,(V\,\Delta z-z\,\Delta V)=\int_{\partial\Omega}\,dS\,
\left( V\,\frac{dz}{dn}-z\,\frac{dV}{dn}\right)&=&\nonumber
\\\int_{S^2(\infty)}\,dS\, \left(
V\,\frac{dz}{dn}-z\,\frac{dV}{dn}\right)-\int_{S}\,dS\,\left(
[V]\,\frac{dz}{dn}-z\,\left[\frac{dV}{dn}\right]\right),\label{Green} 
\end{eqnarray} taking into account that both the cartesian coordinate $z$ and the
potential $V$ are solutions of the Laplace equation in $\Omega$ and that the
boundary $\partial\Omega$ consists of the sphere at infinity and the surface
$S$.

The integral at infinity can be performed if $V$ has the usual multipolar
expresion,

\begin{equation}
V=\frac{Q}{r}+\frac{M\,\cos\theta}{r^2}+O(r^{-3}),
\end{equation}
and it yields the value $4\,\pi\,M$. Hence the integrand on
$S$ can be interpreted as the surface density for the dipole moment $M$ of the
source.

This interpretation arises immediately from potential theory (cfr. for instance
\cite{Kel}), which provides the expressions for the dipole density of a double
distribution,

\begin{equation} 
\sigma_{dip}=\frac{1}{4\,\pi}\,n\cdot u_z\,[V] ,
\end{equation}
in terms of the discontinuity of the potential, and the density of a monopole
layer, (\ref{mono}), related to the jump of the normal derivative of the
potential on the surface. 

Hence the first term in (\ref{clasden}) is the contribution of a dipole
layer to the total dipole density and the second term is just
the moment density in the $z$ direction of the distribution of monopoles. 

\section{Relativistic layers}\label{relat}

For the stationary axisymmetric Einstein-Maxwell system we shall follow a
similar approach, which can be viewed as a generalization of the formalism 
introduced in \cite{second}, \cite{tesis} for static electrovacs. In that paper
magnetic surface-sources were constructed for magnetostatic solutions of the
Einstein-Maxwell equations. The magnetic moment density was calculated from the
discontinuities of the Ernst magnetic potential \cite{Ernst} involved in the
generation of the solutions. However that formalism could not deal with 
asymptotically monopolar electric fields and with nonstatic metrics and,
therefore, many important exact solutions fell out of its scope.

In order to work with the Ernst formalism for the stationary axisymmetric
Einstein-Maxwell system, we shall work in charts adapted to the Killing fields
throughout the paper so that these can be expressed in the form, $
\xi=\partial_t$, $\eta=\partial_\phi$, and restrict ourselves to
metrics which can be rendered into a reducible matrix in a holonomic frame, that
is, there is a set of coordinates in which the metric can be expressed as
$g=g_1\oplus g_2$, where $g_1$ is the metric in the subspace spanned, at each
event, by the Killing fields and $g_2$ is the metric in the subspace  orthogonal
to the former one. Having this restriction in mind, it is useful \cite{ch} to
introduce an orthonormal frame, $\{\theta^0,\theta^1,\theta^2,\theta^3\}$, such
that the Killing fields have non-zero projection on $\theta^0$ and
$\theta^1$ only.

In this orthonormal frame, we shall consider electric and magnetic forms, 
$E$ and $B$, with zero projection on the orbits of the Killing fields, 

\begin{equation}
F=E\wedge\theta^0+*B\wedge\theta^1\ \ \ \ \ \ \ \ \ ^4*F=-B\wedge\theta^0
+*E\wedge\theta^1,
\end{equation}
so that the electromagnetic Faraday, $F$, and Maxwell,$^4*F$, forms have a simple
expression after introducing the Hodge duality ($*\theta^2=\theta^3$) in the
space orthogonal to the Killing fields \cite{second}.

Taking into account Cartan's structure equations, expressing the torsion-free
connection coefficients as one-forms, $a$, $w$, $s$, $b$ and $\nu$, (cfr.
\cite{second},\cite{ch} for the physical meaning of these forms)

\begin{subequations}
\label{cartan}
\begin{equation} 
d\theta^0 = a \wedge \theta^0 - w \wedge \theta^1 
\end{equation} 
\begin{equation}
 d \theta ^1 = (b - a) \wedge \theta ^1 - s\wedge \theta^0
\end{equation}
\begin{equation}
d\theta^2=-\nu\wedge\theta^3 
\end{equation}
\begin{equation}
d\theta^3=\nu\wedge\theta^2 ,
\end{equation}
\end{subequations}

the Maxwell vacuum equations, $dF=0$ and $d\ ^4*F=0$, can be written in a 
compact fashion after defining a complex one-form $f=E+iB$, 

\begin{subequations}
\label{maxwell}
\begin{equation}
df=-a\wedge f-is\wedge *f\label{eq:mrot}
\end{equation}
\begin{equation}
d*f=(a-b)\wedge *f+iw\wedge f\label{eq:mdiv}.
\end{equation}
\end{subequations}

The orthonormal frame can always be chosen without loss of generality so that
the one-form $s$ is zero. Bianchi's compatibility conditions for (\ref{cartan}),

\begin{subequations}
\label{Bianchi}
\begin{equation} 
d b  =  0 \label{eq:cWeyl}
\end{equation} 
\begin{equation} 
d a = w \wedge s \label{eq:acc}
\end{equation}
\begin{equation} 
d w = - (b - 2 a) \wedge w\label{eq:cvor} 
\end{equation} 
\begin{equation} 
d s = (b - 2 a) \wedge s \label{eq:csh},
\end{equation}
\end{subequations}

can be then formally integrated after introducing some new functions $U$, $A$,

\begin{equation}
a=dU\ \ \ \ \ \ \ b=d\ln\rho \label{eq:int}
\end{equation}

\begin{equation}
w=\rho^{-1}e^{2U}dA\label{eq:w}\ \ \ \ \ \ \ s=0,
\end{equation}
which allow the solution of Cartan's first structure exterior system of equations
(\ref{cartan}) and yield the line element in Weyl coordinates,

\begin{equation}
ds^{2}=-e^{2U}(dt-A\,d\phi)^{2}+e^{-2U}\{\rho^2\,d\phi^{2}+e^{2k}(dz^{2}
+d\rho^{2})\}, \label{eq:can} 
\end{equation} 
in terms of the functions $U$, $A$, $k$ of $\rho$ and $z$. 

Also equation (\ref{eq:mrot}) can be solved in terms of a complex scalar
potential, $\Phi$, the Ernst electromagnetic potential \cite{Ernst},

\begin{equation} 
f=E+i\,B=-e^{-U}\,d\Phi\label{eq:f} ,
\end{equation}
that satisfies one of the Ernst equations \cite{Ernst}, 

\begin{equation}
d*d\Phi+(b-2\,a)\wedge
*d\Phi=i\,\frac{e^{-2\,U}}{\rho}\,dA\wedge d\Phi\label{eq:Laplace},
 \end{equation}
which can be easily obtained from equations (\ref{eq:mdiv}) and (\ref{eq:f}).
However, it will be convenient to cast it in a different form, 

\begin{equation}
d(e^{-2\,U}\,\rho*d\Phi-i\,A\,d\Phi)=0,
\end{equation}
for future purposes. In a coordinate patch, it bears a resemblance with a
complex Laplace equation on the hypersurfaces of constant time,

\begin{equation}
L\,\Phi\equiv\frac{1}{\sqrt{g}}\,\partial_\mu\,\left\{N\,\sqrt{g}\,\left(e^{-2\,U}\,
g^{\mu\nu}-\frac{i}{\rho}\,A\,\epsilon^{\mu\nu}\right)\,\partial_\nu\,
\Phi\right\}=0\label{phi}, 
\end{equation} 
but including a correction depending on $A$ due to non-staticity, that prevents
the decoupling of the equations for the real and imaginary parts of $\Phi$. In
this equation $g$ is the metric induced by (\ref{eq:can}) on the hypersurfaces
$t=const.$, $N=(-^4g^{tt})^{-\frac{1}{2}}$ is the lapse function and $\epsilon$
is the Levi-Civit\`a tensor on the surfaces of constant time, $t$, and azimuthal
angle, $\phi$. For simplicity the whole equation has been written as the action
of a differential operator, $L$, on the potential.

As a matter of fact, this is a consequence only of the Maxwell equations in the
curved spacetime whose metric is given by (\ref{eq:can}), regardless of whether
the electromagnetic field is the source of the gravitational field,
since we have not made use of the Einstein equations. Therefore what follows is
valid for any potential that satisfies the equation $L\Phi=0$ on the previously
described geometry.

As it was stated in the first section, we are interested in compact sources
and therefore we shall only consider metrics which can be rendered asymptotically
flat in some coordinates $(t,r,\theta,\phi)$, 

\begin{eqnarray}
ds^2&=&-(1-\frac{2m}{r})\,(dt+\frac{2J\sin^{2}\theta}{r}d\phi)^2+
\nonumber\\&+& (1+\frac{2m}{r})\,\{dr^2+(r^2+c_1\,r)(d\theta^2 +\sin^2\theta
d\phi^2)\}+O(1/r^2) ,
\end{eqnarray} 
that allows us to read the total mass, $m$, and the total angular momentum of
the source, $J$, from the Lense-Thirring expansion. In this set of
coordinates, the electromagnetic potential of the compact
source will be required to have an asymptotic expansion in terms of the
first multipole moments, the total charge $e$ and the complex
electromagnetic dipole $M$, whose real and imaginary parts are, respectively, the
electric and magnetic dipole moment,

\begin{equation}
\Phi=\frac{e}{r}+\frac{M\,\cos\theta}{r^2}+\frac{c_2}{r^2}+O(r^{-3}).
\end{equation}
The constants $c_1$ and $c_2$ may arise in some choices of coordinates.

In the case of a non-smooth or discontinuous electromagnetic Ernst potential,
$\Phi$, the equation $L\Phi=0$ may not hold in the whole manifold, and it may
have a distributional source located in the region where the discontinuity
takes place. Since the aim of this paper is devoted to thin layers in stationary
axisymmetric spacetimes, we shall assume that this region is a closed surface
$S$ of outer unit normal $n$ within the constant time hypersurfaces. We shall
denote these hypersurfaces as $(V_3,g)$ for our calculations.

In order to mimic the construction done in section \ref{potential} it will be
necessary to obtain a Green identity for the $L$ operator. A most natural
candidate can be checked to be,

\begin{eqnarray} \label{green2}
\int_{\Omega}\sqrt{g}\,(Z\,L\Phi-\Phi\,L^+Z)\,dx^1dx^2dx^3=\nonumber\\=
\int_{\partial\Omega}\,dS\,N\left\{e^{-2U}\left(Z\,\frac{d\Phi}{dn}-\Phi\,\frac{dZ}{dn}
\right)+i\,\frac{A}{\rho}\,\left(Z\,*d\Phi(n)+\Phi\,*dZ(n)\right)\right\}, 
\end{eqnarray}
which is a consequence of the divergence theorem. $L^+$ is just the complex
conjugate of $L$. We shall apply this identity to the region $\Omega=V^+_3\cup
V^-_3$, the disjoint union of the outer and inner regions of $V_3$ referred to
the surface $S$. The oriented boundary of $V_3^-$ is just the surface $S$ whereas
that of $V_3^+$ is formed by the sphere at infinity and $S$.

So far no condition has been imposed on the function $Z$. In the classical case,
it was just the cartesian coordinate $z$ and the Laplace equation held for it.
Since we no longer have a Laplace operator but $L$, it seems natural to choose
$Z$ so that it satisfies $L^+Z=0$ in $\Omega$. Hence the left hand side in
(\ref{green2}) will be zero and we shall have to deal with surface integrals
only. In addition to the differential equation  the function $Z$ will be
required to behave near infinity as the cartesian coordinate. The set of
conditions on $Z$ will be then,

\begin{equation} L^+\,Z=0\ \ \ \ \ \ \ \
Z=(r+c_3)\,\cos\theta+O(r^{-1}),\label{zeda}
\end{equation} 
which requires solving an elliptic partial differential equation with
boundary conditions at infinity. Again $c_3$ is a constant.

 The integral at infinity in (\ref{green2}) can be calculated from the
information given by the asymptotic behaviour. We are left with just an
integral over the surface $S$, where the source is located, 

\begin{equation}
0=-4\,\pi\,M+\int_S\,dS\,N\left\{e^{-2U}\left[\Phi\,\frac{dZ}{dn}-Z\,
\frac{d\Phi}{dn}\right]-i\,\frac{A}{\rho}\,\left[Z\,*d\Phi(n)+\Phi\,
*dZ(n)\right]\right\} ,
\end{equation}
which allows us to express the total electromagnetic moment as an integral over
the source,

\begin{equation}
M=\frac{1}{4\,\pi}\,\int_S\,dS\,N\left\{e^{-2U}\left[\Phi\,\frac{dZ}{dn}-Z\,
\frac{d\Phi}{dn}\right]-i\,\frac{A}{\rho}\,\left[Z\,*d\Phi(n)+\Phi\,
*dZ(n)\right]\right\}.
\end{equation}

 As it was done in the classical potential theory, we can interpret the
discontinuity of the integrand as the electromagnetic dipole moment surface
density of the source for the field,

\begin{equation} 
\sigma_M=\frac{1}{4\pi}\,N\left\{e^{-2U}\left[\Phi\,
\frac{dZ}{dn}-Z\,\frac{d\Phi}{dn}\right]
-i\,\frac{A}{\rho}\,\left[Z\,*d\Phi(n)+\Phi\,*dZ(n)\right]\right\},
\label{magden} \end{equation} 
the real part of $\sigma_M$ being the electric dipole density and its
imaginary part, the magnetic moment density.

 When $A$ is zero, that is in the static case, this formula is pretty similar
to the classical one (\ref{clasden}), corrected by metric factors. Its first
term also  coincides in the static case with the formula introduced in
\cite{second}, where the contribution of the moment density arising from
the layer of monopoles (second term) was not taken into account. 

Also the electric charge density on $S$ can be calculated using the
divergence theorem, since equation (\ref{phi}) has the form of a total
derivative and therefore its integral on $\Omega$ can be reduced to a
surface integral on its boundary,

\begin{equation} 
0=\int_{\Omega}\,L\Phi=\int_{\partial
\Omega}\,dS\,N\left\{e^{-2U}\,\frac{d\Phi}{dn}
+i\,\frac{A}{\rho}\,*d\Phi(n)\right\}. \label{fff}
\end{equation}

The asymptotic expansion of the fields provides the necessary information
to perform the integral at infinity,

\begin{equation} 
0=-4\,\pi\,Q-\left\{\int_{S}\,dS\,N\left(e^{-2U}\,\left[\frac{d\Phi}{dn}\right]
+i\,\frac{A}{\rho}\,\left[*d\Phi(n)\right]\right)\right\} ,
\end{equation}
from which we can read the expression for the charge density,

\begin{equation} 
Q=\int_S\,dS\,\sigma_Q\ \ \ \ \ \ \ \ \ \ \sigma_Q=
-\frac{1}{4\pi}\,\Re\left\{N\left(e^{-2U}\,\left[\frac{d\Phi}{dn}\right]
+i\,\frac{A}{\rho}\,\left[*d\Phi(n)\right]\right)\right\}\label{charden} ,
\end{equation}
in terms of the discontinuities of the derivatives of the electromagnetic
Ernst potential $\Phi$. This formula recovers Israel's
expression for the electric charge density \cite {is1}, but written in
terms of the electromagnetic Ernst potential. 

The function $Z$ can only be considered as a coordinate in the static case.
Otherwise the equation (\ref{zeda}) is complex and so is the solution $Z$. This
is very similar to what happens with the Ernst potential \cite{Ernst},
$\varepsilon=e^{2U}+i\chi$. Whereas its first term is a metric function, the
second term is just an auxiliary potential, the twist potential $\chi$, that is
due to nonstaticity. Similarly, just the real part of $Z$ can be viewed as the
coordinate function which determines the projection of the dipole moment that is
being calculated. The imaginary part is again an auxiliary potential which states
the influence of the electric (magnetic) field on the magnetic (electric) dipole
density. 

As it has already been mentioned, this results are valid for any potential
that satisfies the equation (\ref{phi}) in the geometry defined by
(\ref{eq:can}). Therefore the same considerations can be applied to the
gravitational Ernst potential $\varepsilon$ \cite{Ernst}, since it fulfils
$L\,\varepsilon=0$. It will be convenient, however, to introduce another
potential $\eta$,

\begin{equation}
\eta=\varepsilon-1,
\end{equation}
since the Ernst potential is defined up to a constant, which is usually fixed so
that the potential tends to one at infinity and this asymptotic finite value of
$\varepsilon$ is not very convenient when performing integrals on $\Omega$. If
the potential has the following asymptotic behaviour,

\begin{equation}
\eta=-\frac{2\,m}{r}-i\,\frac{2\,J\,\cos\theta}{r^2}+\frac{c_4}{r^2}+O(r^{-3}),
\end{equation} 
where $c_4$ is again a constant, then all the previous calculations can be
repeated just substituting $\Phi$ for $\eta$ to yield expressions for the mass
and angular momentum densities,
\begin{equation}
\sigma_m=\frac{1}{8\pi}\,N\left\{e^{-2U}\,\left[\frac{d\eta}{dn}\right]
+i\,\frac{A}{\rho}\,\left[*d\eta(n)\right]\right\}\label{masden} 
\end{equation}

\begin{equation}
\sigma_J=\frac{1}{8\pi}\,N\left\{e^{-2U}\left[Z\,\frac{d\eta}{dn}-\eta\,
\frac{dZ}{dn}\right]
+i\,\frac{A}{\rho}\,\left[Z\,*d\eta(n)+\eta\,*dZ(n)\right]\right\},
\label{rotden} 
\end{equation}
on the surface $S$.

\section{An example: The Kerr-Newman spacetime}\label{example}

 As an application of this formalism, we shall calculate the sources
for the physical quantities of the Kerr-Newman spacetime \cite{KN}. The
energy-momentum tensor of a source located on the $r=0$ surface of this
spacetime has already been obtained by Israel \cite {is1} using the formalism of
thin layers  \cite{ISR} and afterwards by L\'opez \cite{lop2} using
distributions. We shall focus our attention on the magnetic moment density
since it is not calculated in those references.

Instead of allowing the radial coordinate $r$ to become negative, we shall
restrict its range to positive values, as it is done in \cite{is1}. This amounts
to identifying points on the hypersurface $r=0$ (both sets of coordinates
$(t,\phi,0,\theta)$ and $(t,\phi,0,\pi-\theta)$ represent the same event), as if
we were working with oblate spheroidal coordinates. This identification causes
that certain functions of the collatitude angle $\theta$ will be discontinuous,
such as the cosine, and hence the Ernst potentials, 

\begin{equation} 
\varepsilon=1-\frac{2\,m}{r-i\,a\,\cos\theta}\ \ \ \ \ \ \ \ \
\Phi=\frac{e}{r-i\,a\,\cos\theta} ,
\end{equation}
and their derivatives will
encounter discontinuities on $r=0$, revealing the presence of the source.

 The Kerr-Newman metric, in Boyer-Lindquist coordinates,  

\begin{eqnarray} ds^2&=&-
(1-\frac{2mr-e^2}{r^2+a^2\cos^2\theta})(dt+\frac{(2mr-e^2)a\,\sin^2\theta}{r^2-2mr+e^2+a^2\cos^2\theta}
d\phi)^2+\nonumber\\&+&(1-\frac{2mr-e^2}{r^2+a^2\cos^2\theta})^{-1}\{(r^2-2mr+a^2+e^2)\sin^2\theta
d\phi^2 +\nonumber\\&+&(r^2-2mr+a^2\cos^2\theta+e^2)(
\frac{dr^2}{r^2-2mr+a^2+e^2}+d\theta^2)\}   
\end{eqnarray}
if the parameters $m$, $a$, $e$ satisfy $m^2<a^2+e^2$, induces a line element on
the surface $r=0$, $t=$const.

\begin{equation}
ds^2_2=a^2\,\cos^2\theta\,d\theta^2+\sin^2\theta\,(a^2-e^2\,\tan^2\theta)\,d\phi^2
,\label{disk}\end{equation}
and unit normal,

\begin{equation} 
n=\frac{\sqrt{e^2+a^2}}{a\,\cos\theta}\,\partial_r,
\end{equation}
where $m$ is the total mass, $ma$, the total angular momentum and $e$, the total
electric charge.

It is clear that the Kerr-Newman metric and Ernst potentials fulfil the
previously described asymptotic requirements and therefore the surface densities
can be calculated within this formalism. We need a solution for (\ref{zeda}),

\begin{equation} 
Z=(r-2\,m)\,\cos\theta+\frac
{e^{2}\,\cos\theta+i\,a\,m\cos^2\theta}{r+i\,a\,\cos\theta} ,
\end{equation}  
before calculating the densities on the surface $r=0$ by introducing these
expressions in (\ref{magden}), (\ref{charden}), (\ref{masden}), (\ref{rotden}).

From the expressions for the electromagnetic moment and the charge surface
density,

\begin{subequations}
\begin{equation} 
\sigma_M={\frac {\left(e^{2}\,\cos^2\theta+e^{2}+a^{2}\,\cos^2\theta\right )
\,i\,e}{2\,\pi\,a^{2}\,\cos^3\theta\,|a^{2}-e^{2}\,\tan^2\theta|^{\frac{1}{2}}}}
\ \ \ \ \ \
\theta\in [0,\theta_0)\cup(\theta_0,{\pi}/{2}), 
\end{equation}

\begin{equation} 
\sigma_Q=-{\frac {e}{2\,\pi\,a\,\cos^3\theta\,|a^{2}-e^{2}\,\tan^2\theta
|^{\frac{1}{2}}}}\ \ \ \ \ \
\theta\in [0,\theta_0)\cup(\theta_0,{\pi}/{2}), 
\end{equation}
where $\theta_0=\arctan(|{a}/{e}|)$. 
The corresponding angular momentum and mass densities are obtained by
multiplying them, respectively, by the inverse of the gyromagnetic ratio,
$e/m$, since the Ernst potentials are linearly dependent, 

\begin{equation} 
\sigma_J={\frac {\left(e^{2}\,\cos^2\theta+e^{2}+a^{2}\,\cos^2\theta\right )
\,i\,m}{2\,\pi\,a^{2}\,\cos^3\theta\,|a^{2}-e^{2}\,\tan^2\theta|^{\frac{1}{2}}}}
\ \ \ \ \ \
\theta\in [0,\theta_0)\cup(\theta_0,{\pi}/{2}), 
\end{equation}

\begin{equation} 
\sigma_m=-{\frac {m}{2\,\pi\,a\,\cos^3\theta\,|a^{2}-e^{2}\,\tan^2\theta
|^{\frac{1}{2}}}}\ \ \ \ \ \
\theta\in [0,\theta_0)\cup(\theta_0,{\pi}/{2}). 
\end{equation}

\end{subequations}

The electromagnetic dipole density, $\sigma_M$, is imaginary, and hence there is
no electric dipole density, just the magnetic moment density. 

From the expressions for the differential elements of magnetic moment and electric
charge,

\begin{equation} 
dM_{mag}={\cal I}\{\sigma_M\}\, dS={\frac {\left
(e^{2}\,\cos^2\theta +e^{2}+a^{2}\,\cos^2\theta \right
)\,e\,\sin\theta}{2\,\pi\,a\,\cos^2\theta}} \,d\theta\,d\phi,   
\end{equation}

\begin{equation} 
dQ=\sigma_Q\,dS=-{\frac {e\,\sin\theta}{2\,\pi\,\cos^2\theta }} 
\,d\theta\,d\phi ,  
\end{equation}
we could get by integration on the surface $r=0$ the total magnetic moment and
electric charge of the source. However, instead of obtaining the values
$M_{mag}=e\,a$ and $Q=e$, the integration yields divergent results due to the
singular ring at $r=0$, $\theta=\pi/2$. More precisely the integrals 
\begin{subequations}
\begin{equation}
M=\lim_{\varepsilon\rightarrow
0}\int^{2\,\pi}_0d\phi\int^{\arccos\frac{\varepsilon}{a}}_0d\theta\,
\sigma_M=i\,e\,a \left(1+\lim_{\varepsilon\rightarrow
0}\frac{e^2}{a\,\varepsilon}\right) 
\end{equation}

\begin{equation}
Q=\lim_{\varepsilon\rightarrow
0}\int^{2\,\pi}_0d\phi\int^{\arccos\frac{\varepsilon}{a}}_0d\theta\,
\sigma_Q=e\left( 1-\lim_{\varepsilon\rightarrow 0}\frac{a}{\varepsilon}\right),
\end{equation}
\end{subequations}

provide
the correct results
plus a term that blows up on approaching the singular ring. There is an infinite
contribution to the physical quantities located on the ring.

It is remarkable that the angular momentum density is integrable in the uncharged
case, as it was shown in \cite{first}. From the expression of the induced metric
(\ref{disk}) we learn that there is a change of signature at
$\theta_0$. For values of $\theta$ greater than $\theta_0$, the
metric on the surface is semiriemannian. 

The elements of mass and electric charge are the same that were obtained 
 in \cite{is1} using the thin layer formalism. The linear relation between the
magnetic surface density and the angular momentum surface density confirms the
value of the gyromagnetic ratio, $e/m$, for the Kerr-Newman spacetime. 

There is another choice of location for the source of the Kerr-Newman spacetime,
as it is shown in \cite{lop3}, \cite{gron}. Instead of considering the surface
$r=0$, it is possible to match the Kerr-Newman manifold to Minkowski spacetime
at the pseudosphere $r=r_0=e^2/2m$. The electromagnetic Ernst
potential in the interior of the pseudosphere is taken to be zero. An advantage
of this choice is that the source is valid for all values of the parameters $m$,
$a$, $e$, since this surface is always located in the Boyer-Lindquist chart (the
horizon, whenever there is one, lies within the surface). Another advantage
of this approach lies on the fact that the metric at $r=r_0$,

\begin{equation}
ds^2=-dt^2+(r_0^2+a^2\,\cos^2\theta)\,\left(d\theta^2+\frac{dr^2}{r^2_0+a^2}
\right)+(r_0^2+a^2)\,\sin^2\theta\,d\phi^2,
\end{equation}
is just flat spacetime in oblate spheroidal coordinates, as it happened for the
uncharged case at $r=0$. Therefore the metric tensor is continuous at the
matching surface and also the induced metric, 

\begin{equation}
ds^2_2=(r_0^2+a^2\,\cos^2\theta)\,d\theta^2+(r_0^2+a^2)\,\sin^2\theta\,d\phi^2,
\end{equation}
on the pseudosphere experiences no change of signature.

The expressions that we obtain for the  electromagnetic dipole and monopole
densities on the surface $r=r_0$,

\begin{subequations}
\begin{equation}
\sigma_M=
{\frac {e\,\cos\theta\,\left
\{2\,r_0^2\,(r_0-m)+i\,a\,\cos\theta\,(a^{2}\,\cos^2\theta+3\,r_0^{2})\right
\}\,\sqrt {r_0^{2}+a^{2}}}{4\,\pi \,\left (r_0^{2}+a^{2}\,\cos^2\theta\right )^{5
/2}}}\ \ \ \ \ \ \ \ \ \ \theta\in[0,\pi] \label{nden1} \end{equation}

\begin{equation}
\sigma_Q={\frac {\,e\,\left (r_0^{2}-a^{2}\,\cos^2\theta\right )\,\sqrt
{r_0^{2}+a^{2}} }{4\,\pi \,\left (r_0^ {2}+a^{2}\cos^2\theta\right
)^{5/2}}}\ \ \ \ \ \ \ \ \ \ \theta\in[0,\pi],\label{nden2} \end{equation}
\end{subequations}

lead to the correct results respectively for the total magnetic moment and
electric charge after integration on the pseudosphere. From the fact that
$\sigma_M$ is not imaginary, we learn that there is a non-zero density of
electric moment on the matching surface, given by the real part of
(\ref{nden1}), 

\begin{equation}
\sigma_{M_{elec}}=
{\frac {2\,e\,\cos\theta\,r_0^2\,(r_0-m)\,\sqrt {r_0^{2}+a^{2}}}{4\,\pi
\,\left (r_0^{2}+a^{2}\,\cos^2\theta\right )^{5 /2}}},  
\end{equation}
 although the total electric moment amounts to zero. The magnetic moment
density is then given by the expression

\begin{equation}
\sigma_{M_{mag}}=
{\frac {e\,a\,\cos^2\theta\,(a^{2}\,\cos^2\theta+3\,r_0^{2})\,\sqrt
{r_0^{2}+a^{2}}}{4\,\pi \,\left (r_0^{2}+a^{2}\,\cos^2\theta\right )^{5 /2}}}.
\end{equation}

\section{Discussion}

In this paper a new method for constructing dipole surface sources for
stationary axisymmetric electrovac spacetimes has been introduced. The
expressions that are obtained show the contributions of dipole and monopole
layers to the total dipole density. For this purpose it is necessary to
calculate a function $Z$ as the solution of a linear elliptic differential
equation with boundary conditions at infinity. Its real part can be interpreted
as the function which determines the projection of the dipole  moment that it
is being calculated whereas the imaginary part is an auxiliary potential
related to nonstaticity.

The method has been applied to the Kerr-Newman spacetimes to produce the magnetic
moment density of two different choices of source, both of them on
hypersurfaces of constant time. To our
knowledge, this quantity had not been calculated before. One of them has been
located, for parameters $m$, $a$, $e$ satisfying $m^2<a^2+e^2$, on the
surface $r=0$, that is surrounded by a singular ring,  whereas the other lies on
the pseudosphere $r_0=e^2/2m$. The resulting densities are integrable
in the pseudosphere model but they are not so in the first model. Besides the
pseudosphere model is valid for all ranges of the parameters $m$, $a$, $e$,
since the horizons lie within the surface. Therefore the second choice seems to
be a more suitable source for the Kerr-Newman fields.

For the future it would be interesting to devise new methods for interpreting
the lack of continuity or smoothness of the potentials in terms of distributions
in order to cope with other possible sources, such as struts and rings.

\noindent {\em The present work has been supported in part by DGICYT Project
PB92-0183 and by a FPI Predoctoral Scholarship from Ministerio
de Educaci\'{o}n y Ciencia (Spain). The author wishes to thank F. J. Chinea and 
L. M. Gonz\'alez-Romero for valuable discussions and the referees for their
suggestions.}


\begin{thebibliography}{99}   
\bibitem{Corn} {\it Solutions of Einstein's equations: Techniques and Results} 
(eds.: C. Hoenselaers and W. Dietz, Springer Verlag), Berlin-New York (1984) 
\bibitem{Esc} {\it El Escorial Summer School on Gravitation and General
Relativity 1992: Rotating Objects and Other Topics\/} (eds.: F. J. Chinea and L.
M. Gonz\'alez-Romero), Springer-Verlag, Berlin-New York (1993) 
\bibitem{Ernst} F. J. Ernst, {\it Phys.Rev.} {\bf 168}, 1415 (1968) 
\bibitem{is1} W. Israel, {\it Phys. Rev.} {\bf D2}, 641 (1970)
\bibitem{Kramer} D. Kramer, H. Stephani, M. MacCallum, E. Herlt, {\it Exact
Solution's of Einstein's Field equations} Cambridge University Press,
Cambridge (1980)
\bibitem{Bergh} N. Van den Bergh, P. Wils, {\it Class. Quantum Grav.} {\bf 2},
229 (1984)   
\bibitem{Darmois} G. Darmois {\it M\'emorial des Sciences
Math\'ematiques}  {\bf XXV}, Chap\^{\i}tre 5, Gauthier-Villars, Paris (1927)
\bibitem{ISR} W. Israel, {\it Nuovo Cimento} {\bf 44 B}, 1 (1966) 
\bibitem{Kel} O. D. Kellogg: {\em Foundations of Potential Theory.\/} Dover, New
York, 1954
\bibitem{second} L. Fern\'andez-Jambrina and
F.J. Chinea, {\it Class. Quantum Grav.} {\bf 11}, 1489 (1994) 
\bibitem{tesis} L. Fern\'andez-Jambrina {\it Ph.D. thesis}, Universidad
Complutense de Madrid (1994) (unpublished)
\bibitem{ch}
F. J. Chinea and L.M. Gonz\'alez-Romero, {\it Class. Quantum Grav}. {\bf 9},
1271 (1992) 
\bibitem{KN} E.T. Newman, E. Couch, K. Chinnapared, A. Exton, A. Prakash and R.
Torrence, {\it J. Math. Phys.} {\bf 6}, 918 (1965) 
\bibitem{lop2} C. A. L\'opez, {\it Nuovo Cimento} {\bf 76 B}, 9 (1983)      
\bibitem{first} L. Fern\'andez-Jambrina, F. J. Chinea, {\it Phys. Rev. Lett.}
{\bf 71}, 2521 (1993)
\bibitem{lop3} C. A. L\'opez, {\it Phys. Rev.} {\bf D 30}, 313 (1984)
\bibitem{gron} \O. Gr\o n, {\it Phys. Rev.} {\bf D 32}, 1588 (1985)


\end{thebibliography}
 \end{document}